\pdfoutput=1
\documentclass[iop]{emulateapj}
\usepackage{amsmath,amstext}
\usepackage{mathtools}

\usepackage{bm}
\usepackage{apjfonts}
\usepackage{graphicx}
\usepackage{multirow}
\usepackage{footmisc}

\usepackage[breaklinks,colorlinks,citecolor=blue, linkcolor=blue]{hyperref}
\usepackage{gensymb}

\slugcomment{}

\shorttitle{ISW Effect Measurement Using AllWISE}
\shortauthors{Shajib \& Wright}

\begin{document}
 
\title{Measurement of the integrated Sachs-Wolfe effect using the AllWISE data release}
\author{Anowar J. Shajib\altaffilmark{1}, Edward L. Wright\altaffilmark{1}}
\affil{$^1$Department of Physics and Astronomy, University of California,
Los Angeles, 430 Portola Plaza, Los Angeles, CA 90095; \href{mailto:ajshajib@astro.ucla.edu}{ajshajib@astro.ucla.edu}}

\begin{abstract}
One of the physical features of a dark-energy-dominated universe is the integrated Sachs-Wolfe (ISW) effect on the cosmic microwave background (CMB) radiation, which gives us a direct observational window to detect and study dark energy. The AllWISE data release of the \textit{Wide-field Infrared Survey Explorer} (\textit{WISE}) has a large number of point sources, which span over a wide redshift range including where the ISW effect is maximized. AllWISE data is thus very well-suited for the ISW effect studies. In this study, we cross-correlate AllWISE galaxy and active galactic nucleus (AGN) overdensities with the \textit{Wilkinson Microwave Anisotropy Probe} CMB temperature maps to detect the ISW effect signal. We calibrate the biases for galaxies and AGNs by cross-correlating the galaxy and AGN overdensities with the \textit{Planck} lensing convergence map. We measure the ISW effect signal amplitudes relative to the $\Lambda$CDM expectation of $A=1$ to be $A=1.18 \pm 0.36$ for galaxies and $A=0.64 \pm 0.74$ for AGNs . The detection significances for the ISW effect signal are $3.3\sigma$ and $0.9\sigma$ for galaxies and AGNs respectively giving a combined significance of $3.4\sigma$. Our result is in agreement with the $\Lambda$CDM model.
\end{abstract}


\section{Introduction} \label{sect:intro}
After the discovery of dark energy in the late nineties \citep{Perlmutter98, Riess98}, it became one of the most elusive mysteries in the current-era physics. The existence of dark energy is overwhelmingly, albeit indirectly, evidenced by the measurements of low-redshift Type Ia supernovae, baryon acoustic oscillation, galaxy clustering, and strong lensing \citep[e.g.,][]{Riess09, Reid10, Vikhlinin09, Suyu13}, combined with the measurement of cosmic microwave background (CMB) anisotropies by the \textit{Wilkinson Microwave Anisotropy Probe} (\textit{WMAP}) \citep{Hinshaw12} and \textit{Planck} \citep{Planck15XIII} missions. All these observations suggest our universe to be flat, expanding at an accelerated rate, and dominated by dark energy with approximately 70\% of the energy density of the universe accounted by it.

The integrated Sachs-Wolfe (ISW) effect \citep{Sachs67, Rees68} provides us a method to directly detect the effect of dark energy on CMB photons. When CMB photons cross a gravitational potential well, they experience blueshift while falling in and redshift while going out. The large-scale gravitational potential well is frozen for a matter-dominated, dark-energy-free, flat universe. As a result, the net shift in energy experienced by the CMB photons amounts to zero. However, for a dark-energy-dominated universe the large-scale gravitational potential well decays while the CMB photons are crossing the potential well. Consequently, the photons gain a little amount of energy as the redshift fail to completely compensate the blueshift. This energy shift is approximately one order of magnitude smaller than the primary CMB anisotropies, therefore a direct measurement of the ISW effect is not possible. However, the ISW effect results a correlation between hotter regions in CMB with the large-scale structure (LSS), which can be used as an indirect probe to detect this effect.

Several studies have been performed to detect the ISW effect signal by cross-correlating \textit{WMAP} CMB temperature maps with various survey catalogs and radiation backgrounds, e.g., Sloan Digital Sky Survey (SDSS) luminous red galaxies \citep{Fosalba03, Padmanabhan05, Granett09, Papai11}, 2MASS galaxies \citep{Afshordi04, Rassat07, Francis10}, APM galaxies \citep{Fosalba04}, radio galaxies \citep{Nolta04, Raccanelli08}, and hard X-ray background \citep{Boughn04}. The typical confidence level for the ISW effect detection in the above studies is 2-3$\sigma$. Comprehensive analyses combining different data-sets were carried out  by \citet{Ho08} to detect a 3.5$\sigma$ ISW effect signal and by \citet{Giannantonio08} to achieve the strongest detection to date at 4.5$\sigma$. \citet{Planck15XXI} detected a 4$\sigma$ ISW effect cross-correlation between the \textit{Planck} CMB data and a combination of various data-sets. Using the \textit{Planck} 2015 data release alone, \citet{Cabass15} measured an upper limit for the ISW effect signal amplitude to be $A<1.1$ at 95\% confidence level relative to the $\Lambda$CDM expectation of $A=1$.

The \textit{Wide-field Infrared Survey Explorer} \citep[\textit{WISE};][]{Wright10} conducted an all-sky survey in four mid-infrared frequency bands spanning from 3.4 to 22 $\mu$m. This survey, with millions of galaxies and active galactic nuclei (AGNs),  provides one of the most lucrative data-sets to carry out ISW effect studies. Some earlier studies have been conducted using \textit{WISE} data to detect the ISW effect signal: using \textit{WISE} preliminary release and \textit{WMAP} 7-year data to find a 3.1$\sigma$ detection with the best fit being 2.2$\sigma$ higher than the $\Lambda$CDM prediction \citep{Goto12}; using \textit{WISE} all-sky data and \textit{WMAP} 7-year data to find an 1$\sigma$ detection consistent with the $\Lambda$CDM prediction \citep{Kovacs13}; using \textit{WISE} all-sky data and \textit{WMAP} 9-year data to find a combined 3$\sigma$ ISW effect detection for galaxies and AGNs \citep{Ferraro15}. 

Whereas some of the above mentioned studies reported the signal amplitude of the ISW effect to be in good agreement to the $\Lambda$CDM model \citep[e.g.,][]{Kovacs13, Ferraro15}, some other studies found the ISW effect amplitude to be higher (by 1-2$\sigma$) than that predicted by the $\Lambda$CDM model \citep[e.g.,][]{Ho08, Granett09, Goto12}. \textit{WISE} has detected a large number of point sources over the whole sky and the final AllWISE data release goes roughly twice as deep into the redshift space than the previous all-sky data release according to the AllWISE Explanatory Supplement\footnote{\url{http://wise2.ipac.caltech.edu/docs/release/allwise/expsup/index.html}}. This makes AllWISE data very well-suited to carry out an ISW effect study as the detected objects span a wide range in redshift space that includes where the ISW effect is maximized. In this study, we used the AllWISE and \textit{WMAP} 9-year data-sets to detect the ISW effect signal.

The organization of the paper is as follows. In \autoref{sect:isw}, we briefly review the ISW effect. In \autoref{sect:data}, we describe the data-sets and methods. We present our results in \autoref{sect:results}, followed by discussion and conclusions in \autoref{sect:concl}. Throughout this study, we use \textit{Planck} 2015 results \citep{Planck15XIII}: $H_0 =$ 67.74 $\mathrm{km\ s}^{-1}\mathrm{Mpc}^{-1}$,  $\Omega_m=$ 0.31 and $\Omega_V=$ 0.69 for our fiducial cosmology.

\section{The ISW Effect}\label{sect:isw}

\begin{figure}
\epsscale{1.2}
\plotone{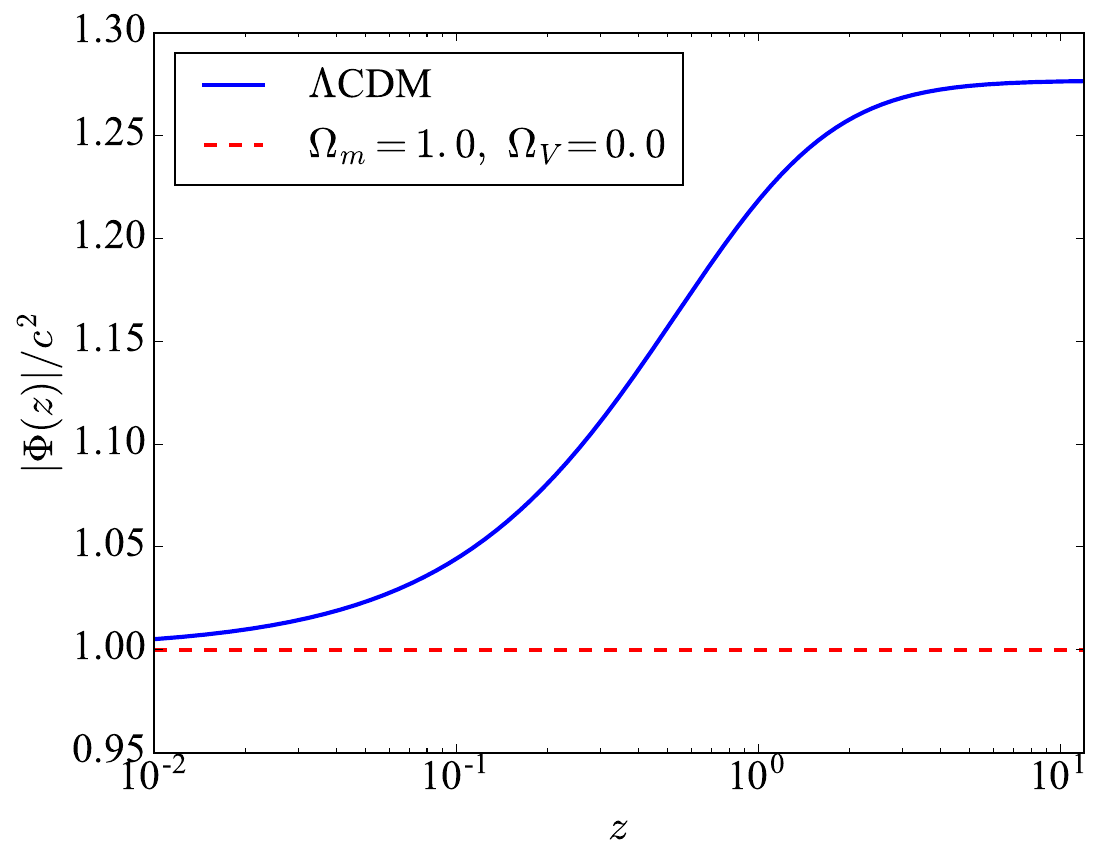}
\caption{
\small
Large-scale gravitational potential as a function of redshift. The potential has been normalized so that $\left| \Phi(0) \right|/c^2 = 1$. Blue solid line is for the $\Lambda$CDM universe with our fiducial cosmology and red dashed line is for a matter-only flat universe. 
\label{fig:phi}
}
\end{figure}

The primary anisotropy in the CMB was created during the last scattering at redshift z$\sim$1100 due to fluctuations of potential energy, photon density, and velocity. The ISW effect is a secondary CMB anisotropy created by the time variation of gravitational potential along the line of sight (\autoref{fig:phi}). This can be expressed as an integral from the last scattering surface to present day as
\begin{equation}
	\begin{aligned}
		\left(\frac{\delta T}{T} \right)_{ISW} (\hat{\bm{n}}) &= - \frac{1}{c^2} \int \left( \dot{\Phi} + \dot{\Psi} \right) \left[ \eta, \hat{\bm{n}} \left( \eta_0 - \eta \right) \right] \\ & \hspace{.1\textwidth} \times e^{-\tau (z)}\ \mathrm{d} \eta\\
		&{} \approx - \frac{2}{c^2} \int \dot{\Phi}\left[ \eta, \hat{\bm{n}} \left( \eta_0 - \eta \right) \right] \ \mathrm{d} \eta ,
		\end{aligned}
	\end{equation}	
where $\eta$ is the conformal time given by $\eta = \int \mathrm{d}t/a(t)$, $a(t)$ is the scale factor, $\dot{\Phi}$ and $\dot{\Psi}$ are the conformal time derivatives of the gravitational potentials $\Phi$ and $\Psi$, $\tau$ is the optical depth, and $e^{-\tau (z)}$ is the visibility function for CMB photons. Here on the second line, we approximated $\tau \ll 1$ over the period when $\dot{\Phi} \neq 0$ to take $e^{-\tau} \approx 1$. We also assumed that anisotropic stresses are negligible, thus we have $\Phi = \Psi$.

As mentioned before, the ISW effect signal is roughly 10 times smaller than the primary CMB anisotropies, thus cleanly separating the ISW effect from the primary anisotropy is not possible. Moreover, the total ISW effect signal includes both positive and negative contributions due to all the small-scale potential fluctuations along the line of sight. We can assume that the ISW effect contributions from the small-scale potential wells and hills cancel each other out within a large enough scale. Then, the significant contribution on the ISW effect signal comes from the LSS. In addition to the ISW effect, the Sunyaev-Zeldovich effect \citep{Sunyaev72} and lensing of CMB photons by matter distribution can also induce a secondary anisotropy that correlates with matter overdensity. However, these anisotropies are only important in small angular scales with multipole $l \gtrsim 100$. We can assume the ISW effect to be the dominant source of secondary anisotropy in the multipole range $l\leq 100$.

To detect the ISW effect signal, we can take a cross-correlation between the CMB temperature anisotropy and the overdensity of a tracer for matter distribution, e.g., galaxies and AGNs. For simplicity we only use subscript or superscript ``$g$'' to denote terms related to the tracer distribution, which are equally applicable for galaxies and AGNs. The tracer overdensity along a given direction $\hat{\bm{n}}$ is given by
\begin{equation}
	\delta_g (\hat{\bm{n}}) = \int b_g(z) \frac{\mathrm{d}N}{\mathrm{d}z} \delta_m(\hat{\bm{n}}, z) \ \mathrm{d} z,
	\end{equation}
where $\mathrm{d}N/\mathrm{d}z$ is the selection function of the survey normalized so that $\int \mathrm{d}N/\mathrm{d}z \ \mathrm{d}z = 1$, $b_g(z)$ is the tracer bias function relating visible matter and dark matter distributions, and $\delta_m$ is the matter density perturbation.

Then, the overdensity-CMB cross-power spectrum is given by
\begin{equation} \label{eq:c_tg}
	C^{Tg}_l = C^{\dot{\Phi}g}_l = 4 \pi T_{CMB} \int \Delta^2_m (k) I^{\dot{\Phi}}_l (k) I^{g}_l (k) \frac{\mathrm{d}k}{k},
	\end{equation}
where $\Delta_m^2 (k)$ is the dimensionless matter power spectrum at redshift $z=0$ given by $\Delta_m^2(k) = k^3 P(k, z=0)/2 \pi^2$ \citep{Cooray02}. The weight functions for the tracer overdensity and the ISW effect are given by
\begin{equation}\label{eq:I_g}
	I^g_l (k) = \int b_g(z) \frac{\mathrm{d}N}{\mathrm{d}z} D(z) j_l \left( k \chi(z) \right) \ \mathrm{d} z,
	 \end{equation}
\begin{equation}\label{eq:I_isw}
	I^{\dot{\Phi}}_l (k) = \frac{3 \Omega_m H_0^2} {c^2 k^2} \int \frac{\mathrm{d}}{\mathrm{d}z} \left[ (1+z) D(z) \right] j_l \left( k \chi(z)\right) \ \mathrm{d} z ,
	\end{equation}
where $j_l$ is the spherical Bessel function, $\chi(z)$ is the comoving distance to redshift $z$ given by $\chi(z) = c\left[\eta_0 - \eta(z)\right]$, and $D(z)$ is the linear growth factor normalized so that $D(z=0) = 1$.

\begin{figure*}
\includegraphics[width=0.5\linewidth]{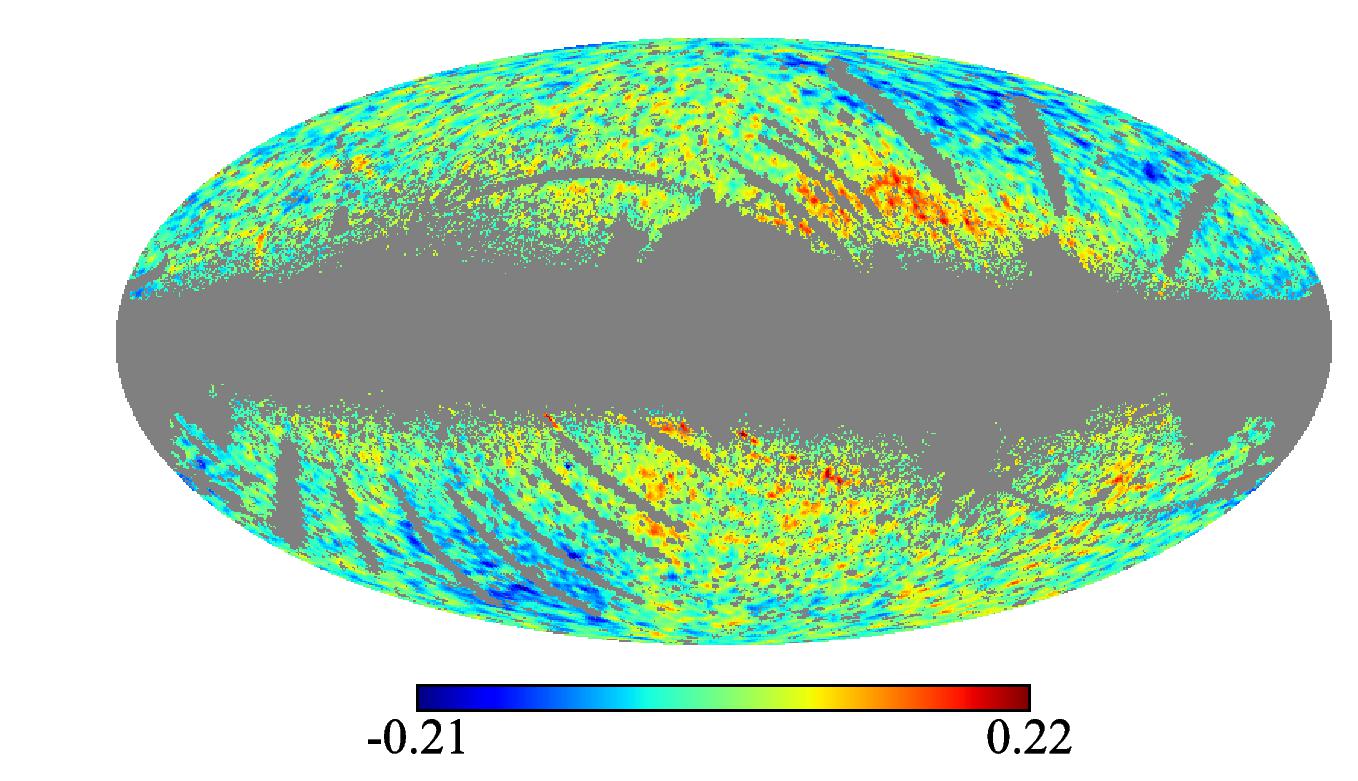}
\includegraphics[width=0.5\linewidth]{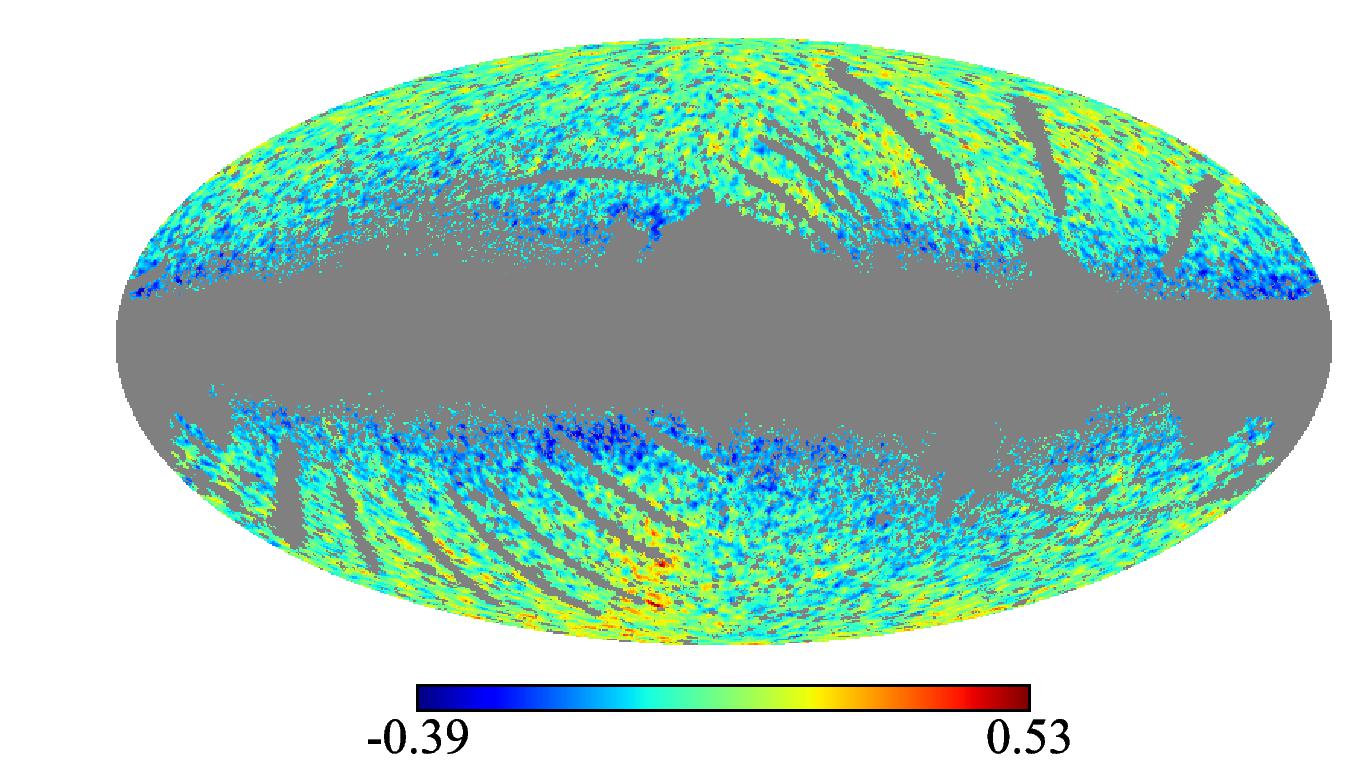}
    \caption{Overdensity maps in galactic coordinate with \textsc{healpix} resolution parameter $n_{side}=128$ for galaxies (left) and AGNs (right). These maps are smoothed with a Gaussian window of standard deviation $\sigma=0.5{\degree}$. The grey area is the mask where the overdensity is zero. The mask leaves the unmasked sky fraction $f_{sky}=0.46$.
		\label{fig:map_overdensity}
		}
\end{figure*}

\newpage
\section{Data and Methods} \label{sect:data}

\subsection{CMB Map}\label{subsect:cmbmap}
We used the 9-year foreground reduced \textit{WMAP} temperature data provided by the LAMBDA website\footnote{\url{http://lambda.gsfc.nasa.gov/}} \citep{Bennett13}. We only used Q, V, and W bands (41, 61, and 94 GHz respectively) as they have the least amount of galactic contamination. As we are only interested in $l \leq 100$, the maps were re-binned into \textsc{healpix} \citep[Hierarchical Equal Area isoLatitude Pixelization;][]{Gorski05} maps with resolution parameter $n_{side} = 128$. We have used the \texttt{KQ75y9} extended temperature analysis mask with $f_{sky} = 0.65$, which excludes point sources detected by \textit{WMAP}. The final mask is the combination of the \textit{WMAP} mask and a mask for the \textit{WISE} data described in \autoref{subsect:mask}. This final mask was applied to both of the maps before taking the cross-correlation.

\subsection{\textit{WISE} Data} \label{subsect:wise}
The \textit{WISE} mission surveyed the whole sky in four bands: 3.4 (W1), 4.6 (W2), 12 (W4), and 22 $\mu$m (W4). In this study, we used the AllWISE data release, which combines the 4-band cryogenic phase with the NEOWISE post-cryo phase \citep{Mainzer11}. This data release is deeper than the previous all-sky data release by roughly a factor of two in W1 and W2 bands as the NEOWISE post-cryo phase only used these two bands. The AllWISE source catalog has over 747 million objects with SNR $\geq 5$ for profile-fit flux measurement in at least one band. We only select sources from the catalog using W1 and W2 magnitudes with SNR $\geq5$ for W1 band and SNR $\geq3$ for W2 band.

The coverage of \textit{WISE} is not uniform throughout the sky. The median number of exposures for the AllWISE data release is $30.17 \pm 0.02$ in W1 and $30.00 \pm 0.03$ in W2 with each exposure being 7.7 s long for both bands. According to the AllWISE Explanatory Supplement, the catalog is 95\% complete for W1 \textless\ 17.1. Therefore, we applied this magnitude cut to ensure uniformity and completeness for our galaxy sample.

In this study, galaxies are defined as sources in the AllWISE catalog that are not classified as stars or AGNs. To remove stars from the object catalog, we used the color cut: [W1-W2 $<0.4$ \& W1 $<10.5$] \citep{Jarrett11}. We also removed any object with W1 - W2 $< 0$ to effectively remove galactic stars \citep{Ferraro15, Goto12}. To select AGNs from the catalog, we used the color cut criterion 
\begin{equation}
	\mathrm{W1} - \mathrm{W2} > 0.662  \exp \left[ 0.232  (\mathrm{W2}-13.97)^2 \right]
\end{equation}
\citep{Assef13}.

For some of the objects in the AllWISE catalog, the W1 source flux uncertainty could not be measured because of the presence of a large number of saturated pixels in 3-band cryo frames containing the source. These sources lie along a narrow strip of ecliptic longitude and they are marked by \textit{null} values for \texttt{w1msigmpro}. These objects are removed from the sample. We also discarded any object with $\texttt{cc\_flags}\neq 0 $ in W1 or W2, as a non-zero value for \texttt{cc\_flags} indicates a spurious detection (diffraction spike, persistence, halo, or optical ghost). After applying the SNR and magnitude cuts, we are left with approximately 383 million objects. Out of these, roughly 192 million (50.0\%) are classified as galaxies, 189 million as stars (49.3\%), and 2.6 million (0.7\%) as AGNs according to the adopted color cut criteria.

\subsection{Mask} \label{subsect:mask}
We constructed the mask for the overdensity-CMB cross-correlation analysis with \textsc{healpix} resolution parameter $n_{side}=128$. The \texttt{moon\_lev} flag in the AllWISE catalog indicates the fraction of frames contaminated by moonlight among the number of frames where the flux from a source was measured. We added \textsc{healpix} pixels with more than 20\% sources with $\texttt{moon\_lev}>2$ to the mask. \textsc{healpix} pixels with more than 10\% sources with $\texttt{cc\_flags}\neq 0$ out of the total source count within the pixel are also added to the mask. As mentioned in \autoref{subsect:wise}, some objects in the AllWISE catalog with \textit{null} values for \texttt{w1msigmpro} were removed from the sample and we excluded regions with more than 1\% of such sources. We also excluded regions with galactic latitude $\left| b \right| < 10^{\circ}$ to effectively remove areas of galactic contamination. For the AGN overdensity map, some \textsc{healpix} pixels ($<$0.2\%) had abnormally high source count and we added these pixels to the mask for the AGN overdensity map. After applying the combined final mask, the unmasked sky fraction becomes $f_{sky} = 0.46$ (\autoref{fig:map_overdensity}). This unmasked region contains approximately 106 million galaxies and 1.5 million AGNs. 

\subsection{Theoretical Computation}
It is computationally difficult to evaluate the spherical Bessel integrals in equations \eqref{eq:I_g} and \eqref{eq:I_isw} through brute force. For efficient computation, we reformulated these integrals as logarithmically discretized Hankel transform following \citet{Hamilton00}. In this form, the integrals can be evaluated through fast-Fourier-transform (FFT) convolutions using the \textsc{FFTLog} algorithm \citep{Talman78}.

Lastly, we used \textsc{camb} with \textsc{halofit} \citep{Lewis00, Smith03} to generate the non-linear matter power spectra for our fiducial cosmology.

\section{Results} \label{sect:results}
\subsection{Redshift Distribution}

\begin{figure}
\includegraphics[width=\linewidth]{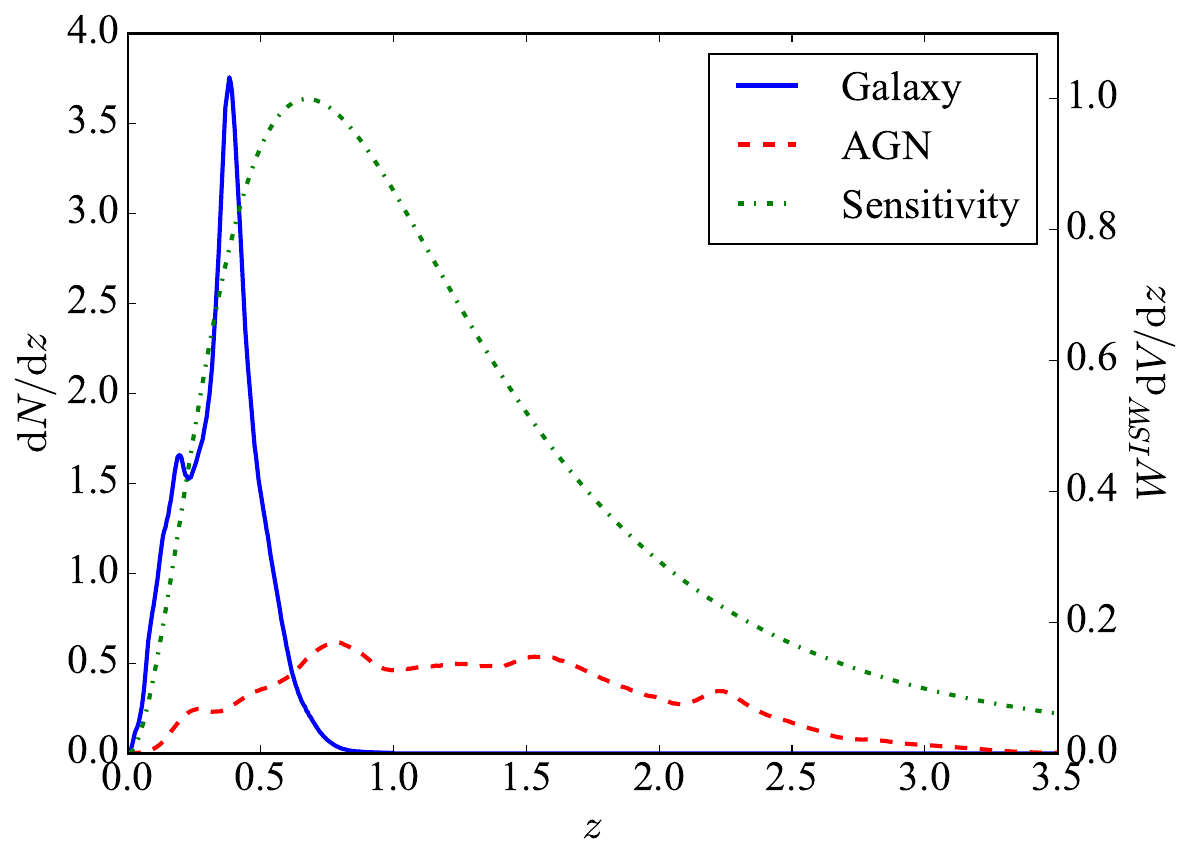}
\caption{
\small
Redshift distribution of AllWISE galaxy (blue solid line) and AGN (red dashed line) samples along with the sensitivity function for the ISW effect cross-correlation (green dotted line). The redshift distribution for galaxies was obtained by cross-matching with SDSS galaxy \texttt{Photoz} catalog and the redshift distribution for AGNs was obtained by cross-matching with SDSS \texttt{SpecObjAll} catalog, with 3$''$ matching radius for both cases. The distributions are normalized so that $\int (\mathrm{d}N/\mathrm{d}z)\ \mathrm{d}z = 1$. The sensitivity function for the ISW effect cross-correlation given by $W^{ISW} \mathrm{d}V/\mathrm{d}z$ is shown with green dashed line, where $W^{ISW} = \mathrm{d}\left[(1+z)D(z)\right]/\mathrm{d}z$ is the ISW effect window function and $V$ is the comoving volume. The sensitivity function is normalized to have a peak value of 1.
\label{fig:photoz}
}
\end{figure}

\begin{figure}
\includegraphics[width=\linewidth]{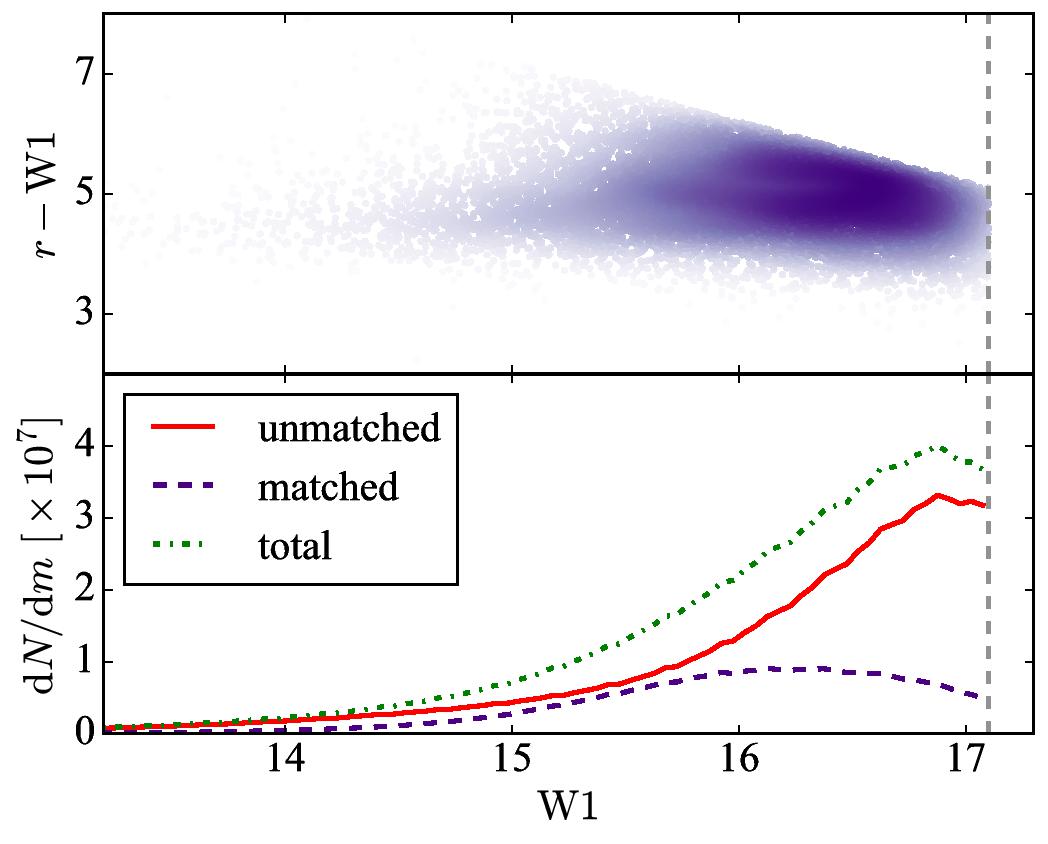}
\caption{
\small
Matching fraction of the \textit{WISE} galaxies with SDSS galaxies for different magnitudes and colors. The top panel shows $r-$W1 vs W1 color distribution of the matched galaxies. Darker area denotes higher density of galaxies and lighter area represents lower density of galaxies in this color-magnitude plot. The bottom panel shows the numbers of SDSS-matched (purple dashed line), unmatched (red solid line), and total (green dotted line) \textit{WISE} galaxies per magnitude bin. The vertical grey dashed line shows the W1$<17.1$ magnitude cut for 95\% completeness. Most of the unmatched galaxies are fainter in W1 and falls behind $r<22.2$ (95\% completeness cut for SDSS).
\label{fig:color}
}
\end{figure}

We performed source matching between SDSS DR12 \citep{SDSS12} galaxy sample and our AllWISE galaxy sample with a matching radius of 3$''$. The matching radius was chosen based on the angular resolutions for \textit{WISE} W1 and W2 bands, which are 6.1$''$ and 6.4$''$ respectively.  We only chose approximately 82 million galaxies with $r>22.2$ \citep[95\% completeness limit, ][]{Abazajian04} from the SDSS DR12 \texttt{Photoz} catalog.\footnote{RA and dec for corresponding sources in the \texttt{Photoz} table are taken from the \texttt{GalaxyTag} table.} The common sky fraction for our mask and SDSS coverage region is $f_{sky} = 0.24$ and it contains approximately 56 million AllWISE galaxies. We find matching pairs for roughly 29\% of the AllWISE galaxy sample. The redshift distribution was then inferred from the SDSS photometric redshift of the matched galaxies (\autoref{fig:photoz}). The low matching percentage of the AllWISE galaxies with SDSS is expected, because high redshift galaxies are optically fainter with redder $r-$W1 color and the majority of the unmatched AllWISE galaxies can be massive ellipticals at $z\gtrsim 1$ \citep{Yan13}. As the 95\% completeness magnitude limit for \textit{WISE}, W1$<17.1$, goes quite deep in the redshift space, many high redshift \textit{WISE} selected galaxies fall beyond the SDSS 95\% completeness limit of $r<22.2$ (\autoref{fig:color}).

To obtain the redshift distribution of the AGN sample, we executed source matching with approximately 750 thousand objects flagged as `QSO' in the SDSS DR12 \texttt{SpecObjAll} catalog, which has spectroscopic redshifts for roughly 4.4 million objects. The matching radius was also taken as 3$''$. Out of roughly 848,000 \textit{WISE} selected AGNs in the common coverage region, we found matching pairs for approximately 15\% of them.

It should be noted that SDSS had an uneven target selection strategy over different redshifts leading to a bias in the redshift distribution of the SDSS objects. Therefore, the redshift distribution obtained by source matching with SDSS objects would also be similarly biased. However, the ISW effect sensitivity function is widespread over a broad range of redshift peaking at $z_{peak}\approx 0.66$ (\autoref{fig:photoz}) and the ISW effect measurement by cross-correlation is not largely sensitive to errors in the estimation of redshift distribution \citep{Afshordi04b}.

\subsection{Bias Measurement}
\begin{figure*}
\includegraphics[width=0.5\linewidth]{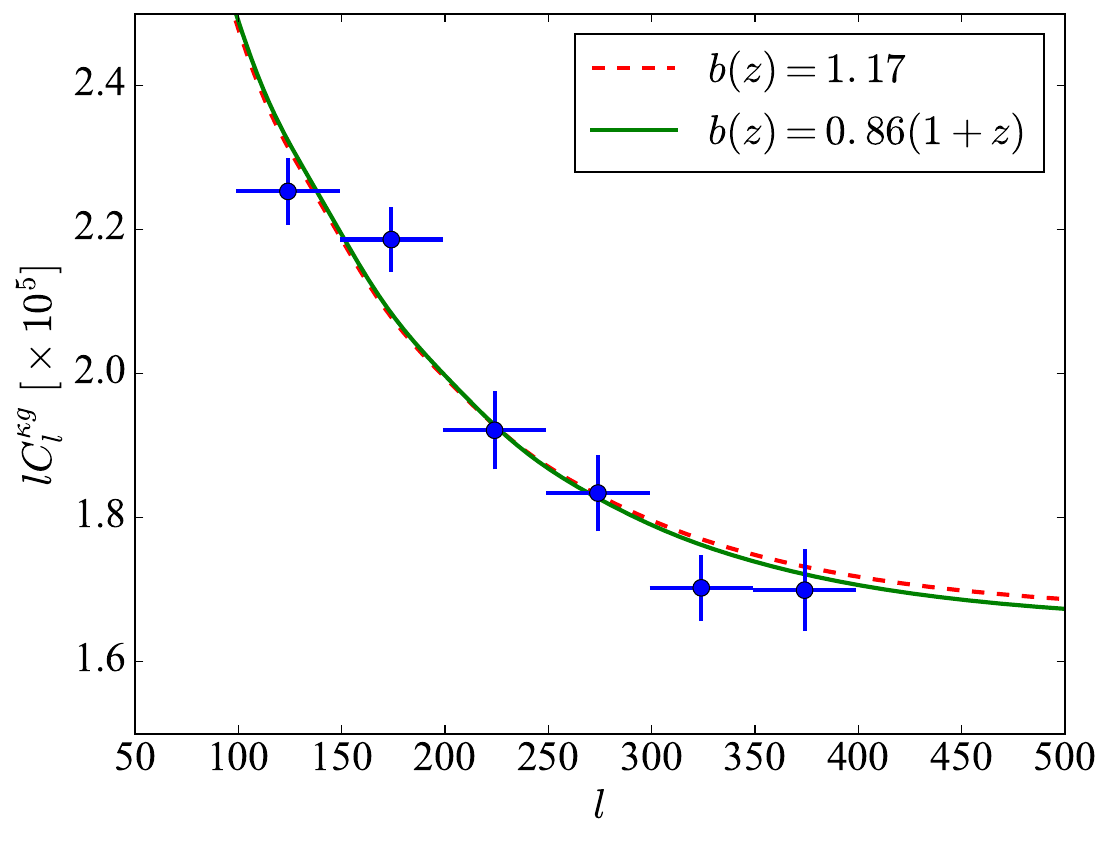}
\includegraphics[width=0.5\linewidth]{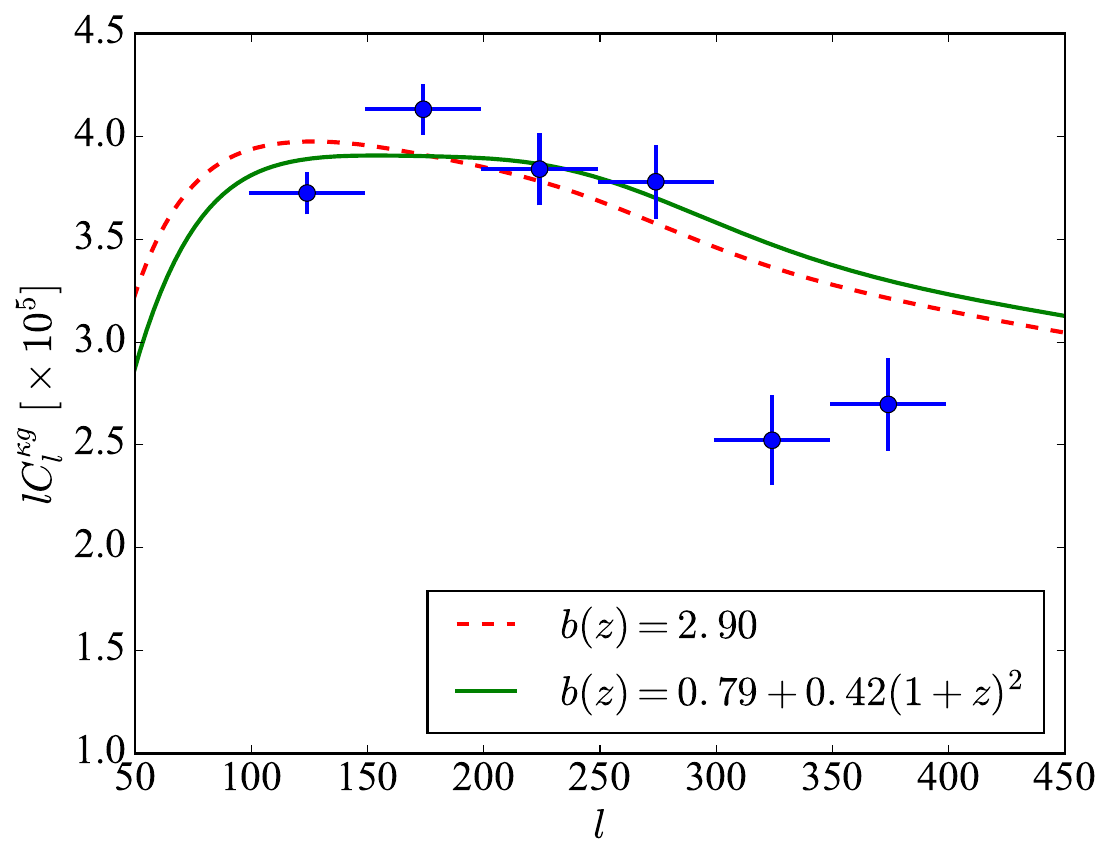}
    \caption{Cross-correlation between \textit{Planck} lensing convergence and \textit{WISE} galaxies (left) and  AGNs (right). Vertical error bars are obtained from 100 simulated lensing convergence maps provided in the \textit{Planck} lensing package and the horizontal error bars show bin widths for the bandpowers. The different bias models used for fitting are shown using lines and described in the corresponding legends. See \autoref{table:bias-fit} for the errors of the best fit parameters for different models.
		\label{fig:bias} 
		}
\end{figure*}

Following \citet{Ferraro15}, we used weak lensing of CMB by our tracers of matter overdensity to measure the bias. This method has two advantages over measuring the bias from auto-correlation of the tracers: (1) it takes into account contamination by stars or artifacts, (2) it is less prone to systematic errors giving a more robust estimation of the bias. The observed CMB temperature $T(\hat{\bm{n}})$ is the lensed remapping of the original CMB temperature field $T_0(\hat{\bm{n}} + \bm{d}) = T(\hat{\bm{n}})$, where $\bm{d}$ is the deflection field. Then, CMB lensing convergence is defined as $\kappa \equiv - \bm{\nabla} \cdot \bm{d} / 2 = - \nabla^2 \phi / 2$, where $\phi$ is the lensing potential. The lensing convergence can be expressed as the line-of-sight integral of matter fluctuation as
\begin{equation}
	\kappa (\hat{\bm{n}}) = \int \delta \left(\chi \hat{\bm{n}},\ z(\chi)\right)\ W^{\kappa} (\chi)\ \mathrm{d} \chi ,
	\end{equation}
where $W^{\kappa}$ is the lensing window function given by
\begin{equation}
	W^{\kappa} (\chi) = \frac{3 \Omega_m H_0^2}{2 c^2} \frac{\chi}{a(\chi)} \frac{\chi_{ls} - \chi}{\chi_{ls}}
	\end{equation}
\citep{Cooray00}. Here, $a(\chi)$ is the scale factor and $\chi_{ls} \approx 14$ Gpc is the comoving distance to the last-scattering surface. 

The cross-correlation between the lensing convergence and matter overdensity field can be calculated using the Limber approximation \citep{Limber53, Kaiser92}, which works well for our angular scale of interest $l \gtrsim 100$, as
\begin{equation}
	C^{\kappa g}_l \approx \int \frac{1}{\chi^2} W^{\kappa} (\chi) W^{g} (\chi) P\left( k = \frac{l+1/2}{\chi}, z \right) \frac{\mathrm{d} \chi}{\mathrm{d} z}\ \mathrm{d}z ,
\end{equation}
where $P(k, z)$ is the non-linear matter power spectrum at redshift $z$ for our fiducial cosmology and $W^g$ is the tracer distribution window function given by
\begin{equation}
	W^g (\chi) = \frac{\mathrm{d}z}{\mathrm{d}\chi} \frac{\mathrm{d}N}{\mathrm{d}z} b(\chi) .
	\end{equation}

We used the lensing convergence map provided by \textit{Planck} data release\footnote{\url{https://irsa.ipac.caltech.edu/Missions/planck.html}} 2 \citep{Planck15XV} to cross-correlate it with the overdensity maps of our LSS tracers to measure their effective biases. The correlation between \textit{WISE} and \textit{Planck} lensing convergence was investigated by \citet{Geach13} and \citet{Planck13XVII}, where these authors found $\sim 7 \sigma$ detection for both galaxies and AGNs. Here, we repeated a similar analysis. We converted the lensing convergence and overdensity maps to \textsc{healpix} resolution $n_{side} = 512$. The mask for this analysis was taken to be a combination of the mask for the ISW effect analysis and the lensing convergence mask provided in the \textit{Planck} lensing package. The unmasked sky fraction for this combined mask is $f_{sky}=0.45$. We obtained the pseudo-power spectrum $\tilde{C}_l^{\kappa g}$ of lensing-overdensity cross-correlation using the \textsc{anafast} facility of the \textsc{healpix} package. We deconvolved the effect of masking and pixelization using the \textsc{master} approach \citep{Hivon02} as
\begin{equation} \label{eq:C^k_l}
	C_{l^\prime}^{\kappa g} = \frac{1}{B_{l^\prime}} \sum_{l} M_{l l^\prime}^{-1} \tilde{C}_{l}^{\kappa g} ,
	\end{equation}
where $M_{ll^\prime}$ is the mode-mode coupling kernel for the applied mask and $B_l$ is the pixel window function for $n_{side} = 512$. We binned the power spectra into six bins (bandpowers) in the multipole range $ 100 \leq l \leq 400$ as $\mathcal{C}_b^{\kappa g} = \sum_{l} P_{bl} C_l^{\kappa g}$, where $P_{bl}$ is the binning operator
\begin{equation}
	P_{bl} = 
	\begin{dcases} 
	\frac{l}{ l^{(b+1)}_{low} - l^{(b)}_{low}}, & \quad \text{if} \quad l^{(b)}_{low} \leq l < l^{(b+1)}_{low}, \\
	0, & \quad \text{otherwise}. \\
	\end{dcases}
	\end{equation}
Here, $l^{(b)}_{low}$ is the lower boundary of the $b$-th bin.

We used 100 simulated lensing convergence maps provided in the \textit{Planck} lensing package to calculate the covariance matrix $\mathrm{C}$ as
\begin{equation} \label{eq:covmat}
	\mathrm{C}_{bb^\prime} = \left\langle (\mathcal{C}_b^{\kappa g} - \left\langle \mathcal{C}_b^{\kappa g} \right\rangle_{sim}) (\mathcal{C}_{b^{\prime}}^{\kappa g} - \left\langle \mathcal{C}_b^{\kappa g} \right\rangle_{sim}) \right\rangle_{sim} ,
\end{equation}
where $\langle\ \rangle_{sim}$ denotes an average over the simulated maps. 

We fit the estimated cross-correlation bandpowers to different bias models for both galaxies and AGNs. Several bias models have been proposed in the literature, e.g., constant bias model $b(z) = b_0$ \citep{Peacock94}, linear redshift evolution model $b(z) = b_0(1+z)$ \citep{Ferraro15}, fitting function for AGNs $b(z) = b_0(0.55+0.289(1+z)^2)$ \citep{Croom05} etc. We fit for the constant and linear evolution bias models in the lensing-overdensity cross-correlation analysis for galaxies, and the constant and fitting function bias models in the lensing-overdensity cross-correlation analysis for AGNs (\autoref{fig:bias}). 

We obtained the best fit for each model by maximizing the likelihood function
\begin{equation}
	\mathcal{L} (\bm{d}; \bm{t}, \mathrm{C}) \propto \exp \left[ -\frac{1}{2} (\bm{d} - \bm{t})^T \mathrm{C}^{-1} (\bm{d} - \bm{t})\right] , 
	\end{equation}
	where $\bm{d}$ is the vector containing measured bandpowers, $\bm{t}$ is the vector containing expected bandpowers of the cross-correlation for each bias model, which depend on the model parameters, and $\mathrm{C}$ is the covariance matrix. Here, we have assumed that individual data points are Gaussian distributed. Maximizing the likelihood function is equivalent to minimizing  $\chi^2 = (\bm{d} - \bm{t})^T \mathrm{C}^{-1}(\bm{d} - \bm{t})$ and the likelihood ratio between two models are given by $-2 \ln (\mathcal{L}_1/\mathcal{L}_2) = \Delta \chi^2$. The best fit parameters for each model are given in \autoref{table:bias-fit}. We used the best fit bias models, linear evolution model for galaxies and constant bias model for AGNs, in the CMB temperature-overdensity cross-correlation analysis.

\begin{deluxetable}{cccc}
	\tablewidth{0pc}
	\tablecolumns{4}
	\tabletypesize{\footnotesize}
	\tablecaption{Best fit parameters for different bias models \label{table:bias-fit}}
	\tablehead{
		\colhead{LSS tracer} &
		\colhead{bias model $b(z)$} & 
		\colhead{$b_0$} & 
		\colhead{$\chi^2$} 
		} 
		
\startdata
\multirow{ 2}{*}{Galaxy sample}   &$b_0$ & 1.17$\pm$0.02\tablenotemark{a} & 10.6  \\
	& $b_0 (1+z)$ & 0.86$\pm$0.01\tablenotemark{a} & 9.7  \\ \\
 
\multirow{ 2}{*}{AGN sample} &  $b_0$ & 2.90$\pm$0.07\tablenotemark{a} & 36.3  \\
	 & $b_0 \left(0.55 + 0.289(1+z)^2\right)$ & 1.44$\pm$0.04\tablenotemark{a} & 37.0 
\enddata

\tablenotetext{a}{The errors are computed by fitting the likelihood function $\mathcal{L}(\bm{d}; \bm{t}(b_0), \mathrm{C}) \propto \exp \left[ (\bm{d}-\bm{t})^T \mathrm{C}^{-1} (\bm{d}-\bm{t})\right] $ to a Gaussian distribution and taking the standard deviation $\sigma$ of the fit as the error. Here, $\bm{d}$ is the vector containing measured bandpowers, $\bm{t}$ is the vector containing expected bandpowers for a given bias model, and $\mathrm{C}$ is the covariance matrix.}

\end{deluxetable}

\subsection{Cross-correlation Measurement} \label{subsect:corrmeasure}
We measured the cross-correlation of \textit{WMAP} CMB maps in Q, V, and W bands and the AllWISE galaxy and AGN overdensity maps. The complex geometry of the mask induces off diagonal correlations between the multipoles. We deconvolved the effect of masking and pixelization from the pseudo-power spectrum $\tilde{C}^{Tg}_l$, which is obtained through \textsc{anafast}, as
\begin{equation}
	C^{Tg}_{l^\prime} = \frac{1}{B_{l^\prime} F_{l^\prime}}\sum_{l} M_{l l^\prime}^{-1} \tilde{C}^{Tg}_{l} , 
	\end{equation}
where $M_{l l^\prime}$ is the mode-mode coupling kernel for our adopted mask, $B_l$ is the pixel window function for $n_{side}=128$, and $F_l$ is the \textit{WMAP} beam transfer function. \textit{WMAP} provides beam transfer functions for each differencing assembly in a band. We took an average of the beam transfer functions for all the differencing assemblies in a given band to obtain the beam transfer function for each band as $F_l^2 = \sum_i^N ( F_l^{(i)}){}^2/N$, where $N$ is the number of differencing assemblies in each \textit{WMAP} band and the index $i$ goes over all the differencing assemblies. We binned the deconvolved power spectra into eight logarithmic bins (bandpowers) using a binning operator $P_{bl}$ given by
\begin{equation}
	P_{bl} = 
	\begin{dcases} 
	\frac{1 }{ 2 \pi} \frac{l(l+1)}{ l^{(b+1)}_{low} - l^{(b)}_{low}}, & \quad \text{if} \quad l^{(b)}_{low} \leq l < l^{(b+1)}_{low}, \\
	0, & \quad \text{otherwise}, \\
	\end{dcases}
	\end{equation}
where $l^{(b)}_{low}$ denotes the lower boundary of the $b$-th bin. We took the bin boundaries as $l=$ 2, 5, 8, 12, 17, 26, 41, 64, 100; thus the first band includes $l=$2, 3, 4 etc. We avoided $l \geq 100$ as the ISW effect is not sensitive to these small scales.

\begin{figure}
	\centering
\includegraphics[width=\linewidth]{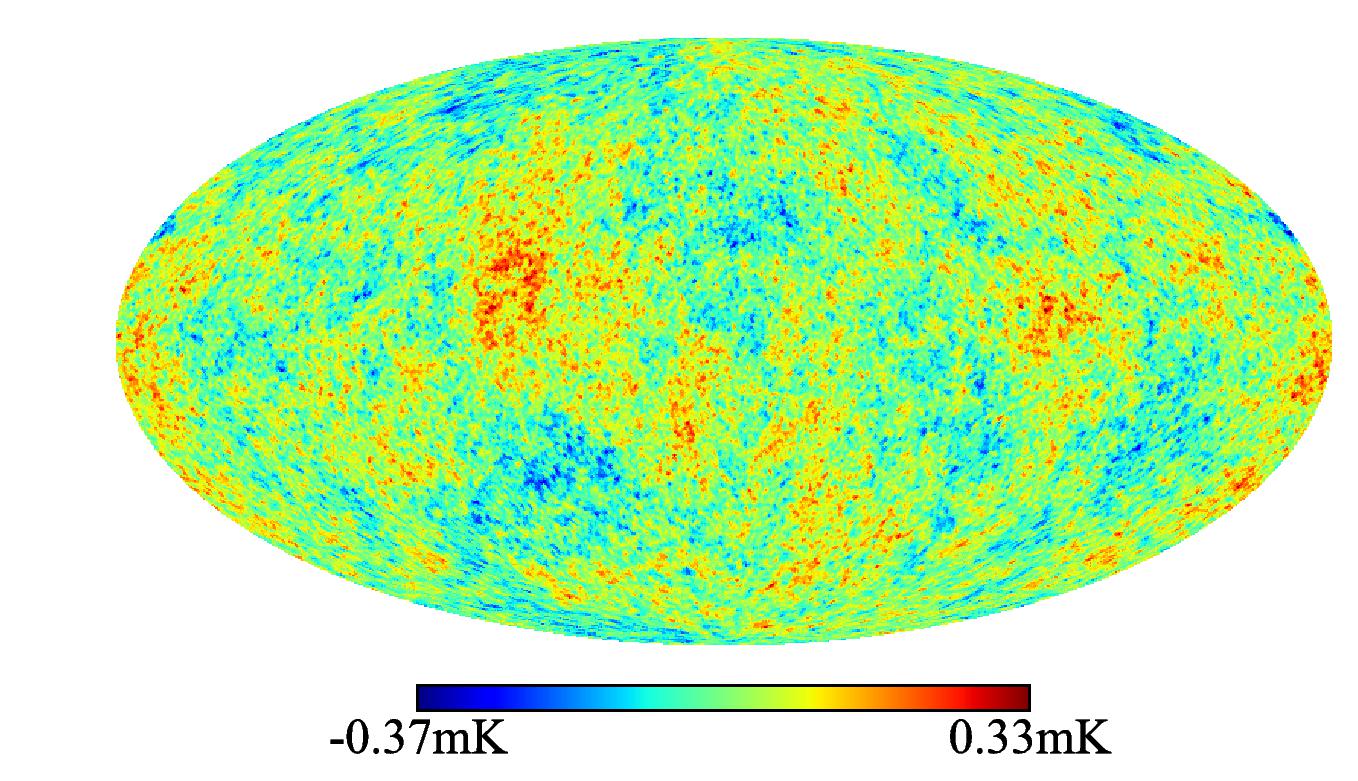}
\includegraphics[width=\linewidth]{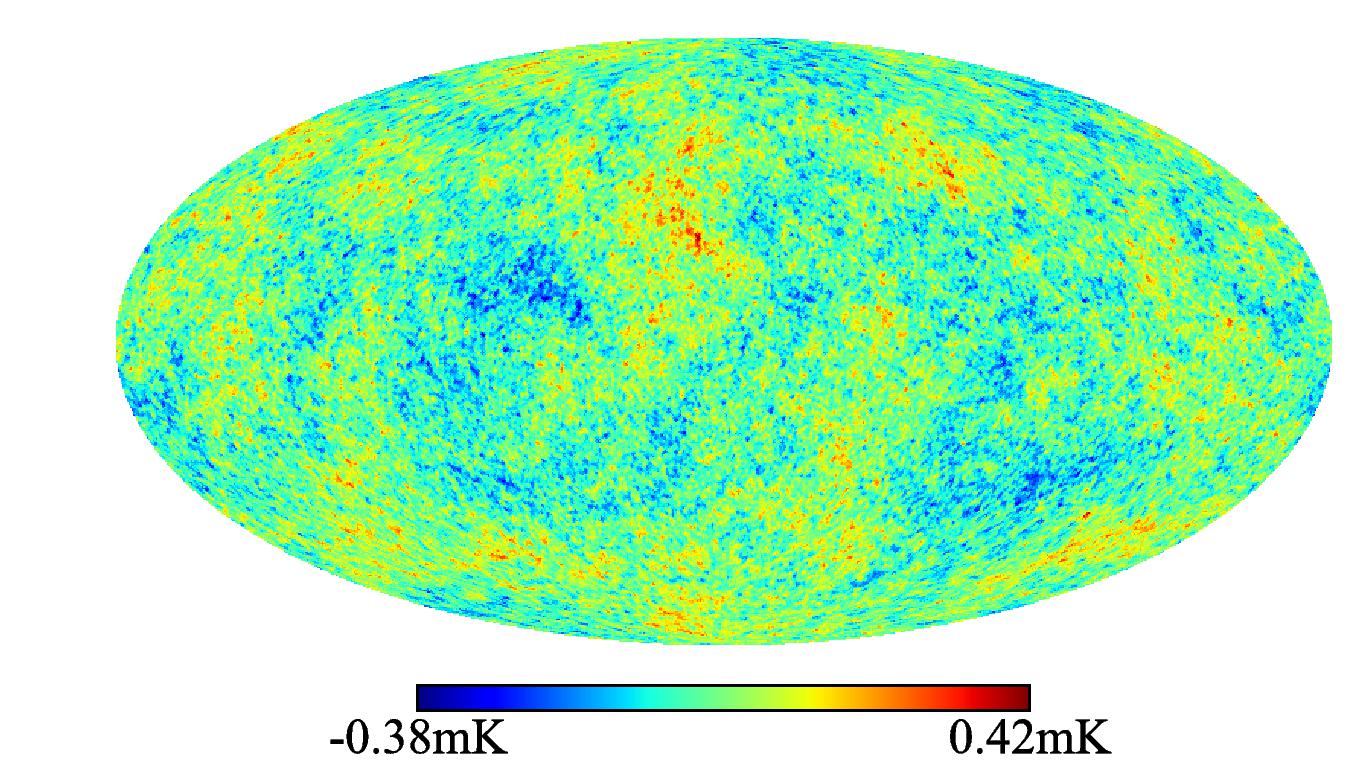}
\includegraphics[width=\linewidth]{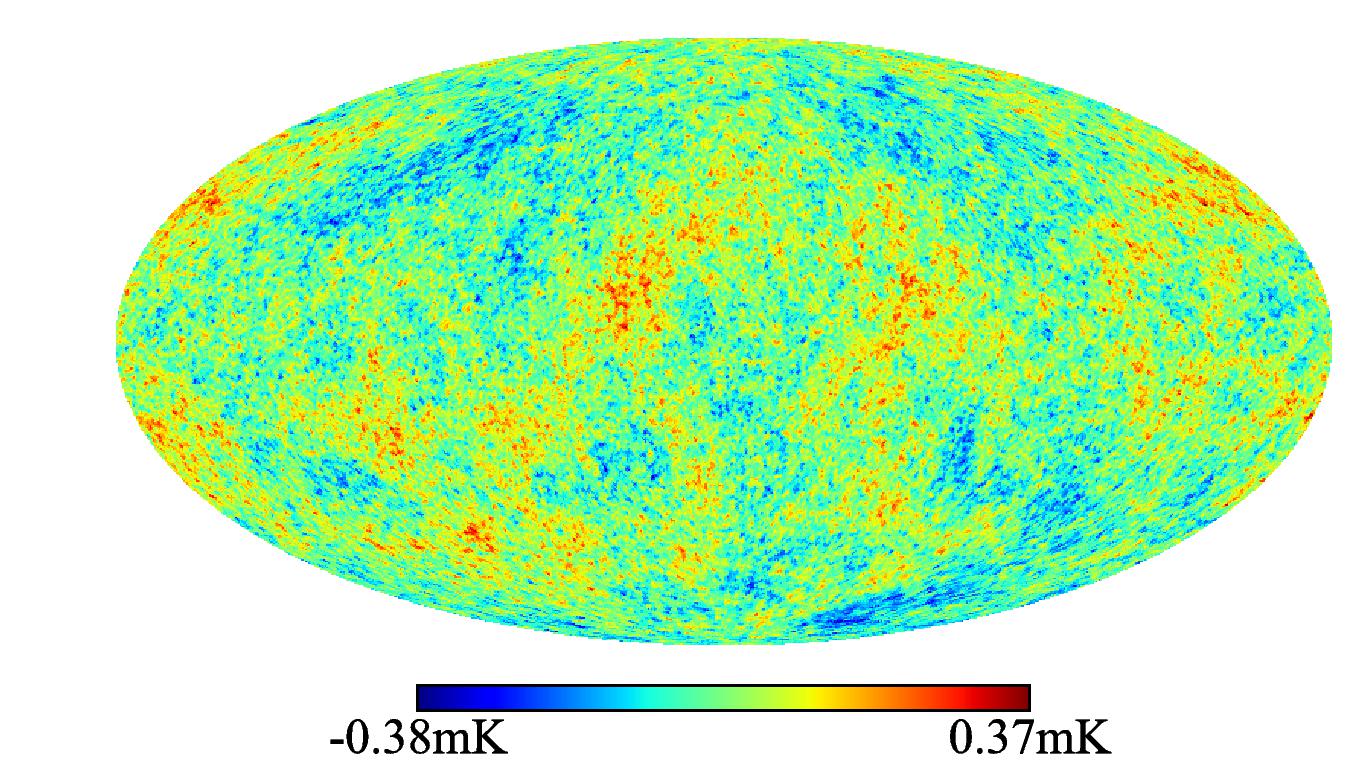}
    \caption{Examples of simulated \textit{WMAP} CMB maps in Q band (top), V band (middle), and W band (bottom) using our fiducial cosmology and \textit{WMAP} beam transfer function. They used different random $a_{lm}$'s, but the same $C_l$ generated using \textsc{camb} for our fiducial cosmology.
		\label{fig:cmbsim}
		}
\end{figure}

\begin{figure}
\includegraphics[width=\linewidth]{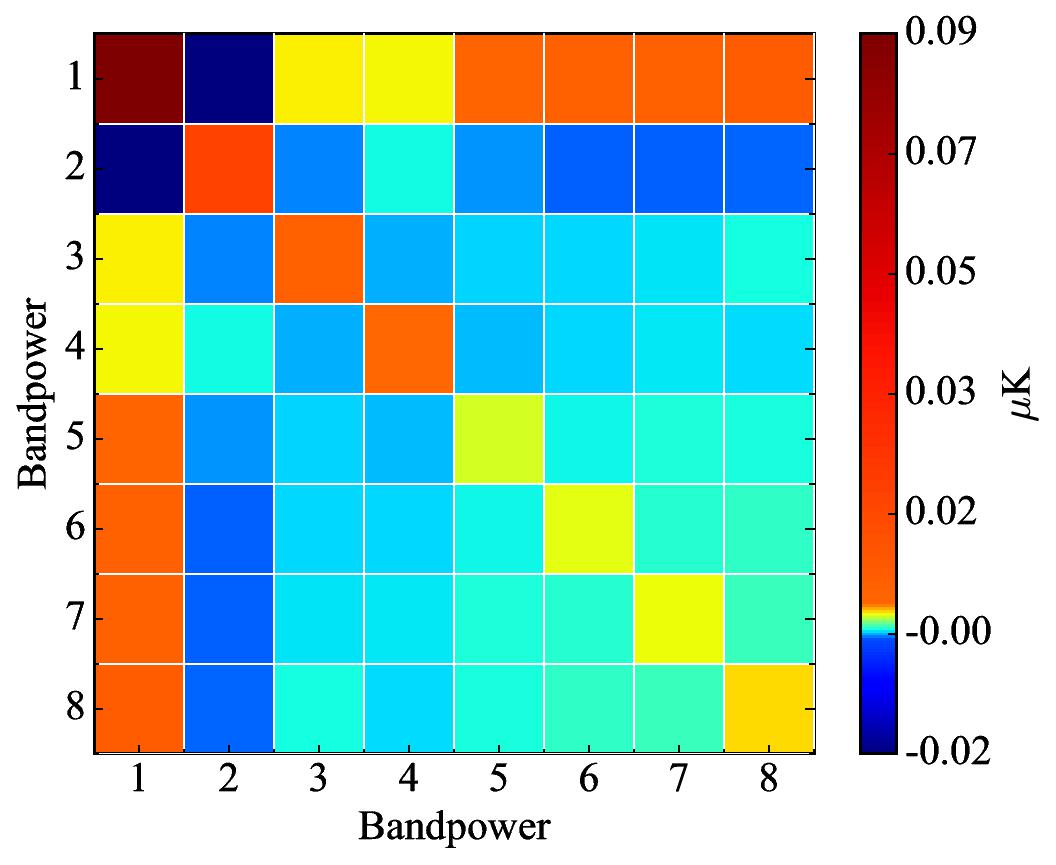}
\includegraphics[width=\linewidth]{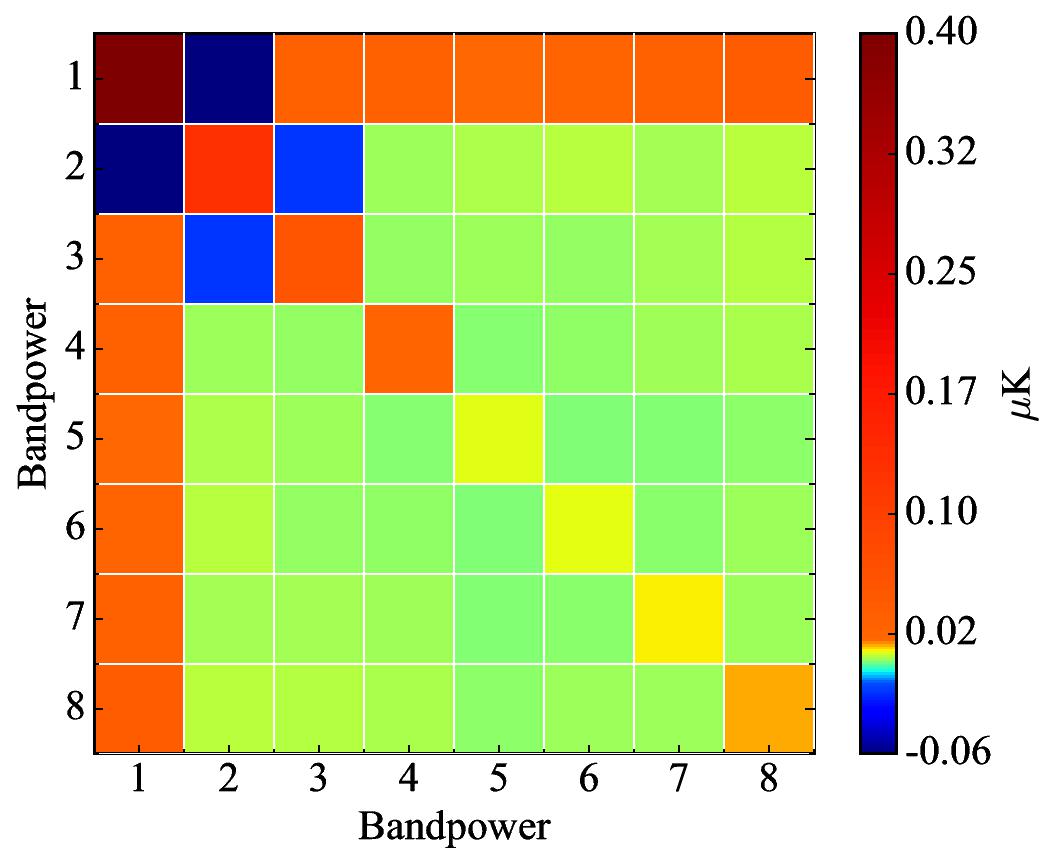}
    \caption{Monte Carlo covariance matrices for galaxy-CMB (top) and AGN-CMB (bottom) cross-correlation bandpowers in Q band. Covariance matrices for V and W bands are not included as they are similar.}
		\label{fig:covmat}
\end{figure}

To estimate the Monte Carlo error bars and covariance matrices, we ran 5000 simulations for each \textit{WMAP} band. We used our fiducial cosmological parameters and \textit{WMAP} beam transfer function to obtain the simulated CMB maps using \textsc{synfast} included in the \textsc{healpix} package. Then, we added noise to each pixel by adding a random value from a Gaussian distribution with zero mean and standard deviation given by $\sigma = \sigma_0 / \sqrt{N_{obs}}$, where $\sigma_0$ is 2.188, 3.131, and 6.544 mK for Q, V, and W bands respectively and $N_{obs}$ is the effective number of observations for the corresponding pixel in the \textit{WMAP} survey. Some examples of the simulated CMB maps for different \textit{WMAP} bands are shown in \autoref{fig:cmbsim}. We cross-correlated these simulated maps with AllWISE galaxy and AGN overdensity maps to obtain the covariance matrices according to equation \eqref{eq:covmat}. The error bars are taken to be the square roots of diagonal elements of the covariance matrix. The neighboring bins are anti-correlated by 3-20\% in the lower multipole range and correlated by roughly 20-30\% in the higher end of the multipole range (\autoref{fig:covmat}).

\begin{figure}
\includegraphics[width=\linewidth]{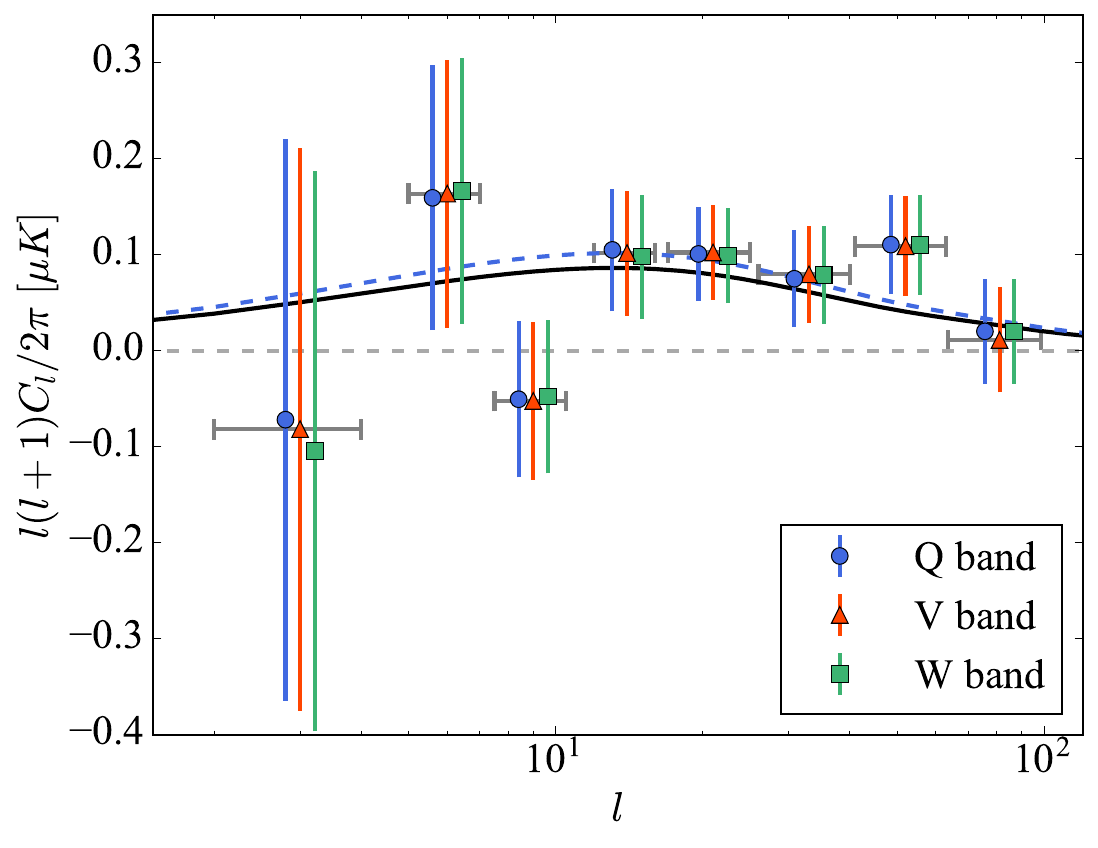}
\includegraphics[width=\linewidth]{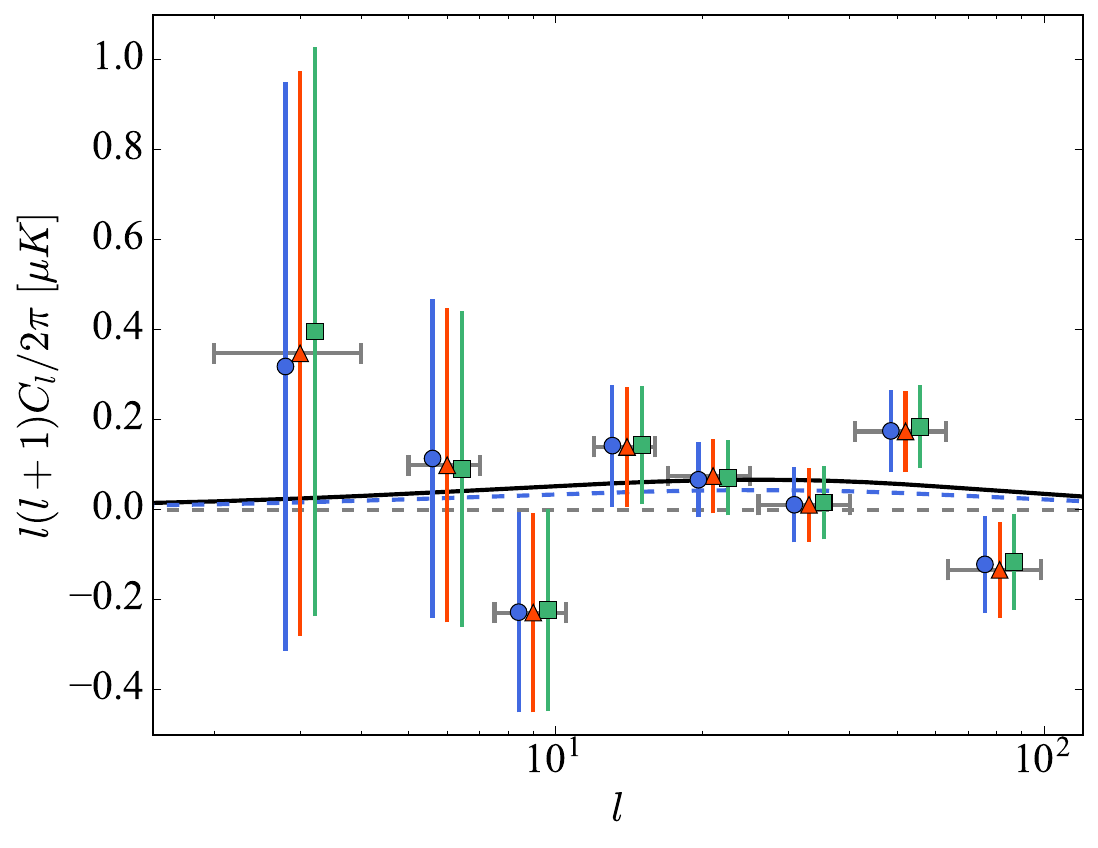}
    \caption{Power spectra of the cross-correlation between \textit{WMAP} CMB maps and AllWISE overdensity maps for galaxies (top) and AGNs (bottom). The power spectra for $2\leq l \leq 100$ in three \textit{WMAP} bands, Q(blue circle), V(red triangle), and W(green rectangle), are binned into eight logarithmic bins. The points for Q and W bands are slightly shifted negatively and positively along $l$ axis for better visual clarity. The vertical error bars are Monte Carlo error bars computed using 5000 simulated CMB maps for each \textit{WMAP} band. The grey horizontal error bars for each group of points show the bin widths. The black solid line shows the theoretical prediction from the $\Lambda$CDM model and the blue dashed line is the best fit for Q band. For both galaxies and AGNs, the measured cross-correlation amplitude agrees very well with the $\Lambda$CDM prediction.}
		\label{fig:cl} 
\end{figure}

We find that the band powers are consistent across different \textit{WMAP} bands (\autoref{fig:cl}). This indicates that the CMB maps are not likely to have significant foreground contamination.

	\begin{deluxetable*}{cccccccc}
		\tablewidth{0.6\linewidth}
		\tablecolumns{8}
		\tabletypesize{\footnotesize}
		\tablecaption{Statistical properties of \textit{WISE}-CMB cross-correlation amplitudes \label{table:gal-isw}}
		\tablehead{
			\colhead{LSS tracer} & 
			\colhead{\textit{WMAP} Band} & 
			\colhead{$A$} & 
			\colhead{S/N} & 
			\colhead{$\chi^2$} & 
			\colhead{d.o.f.} & 
			\colhead{$\Delta \chi^2_{\Lambda CDM}$} & 
			\colhead{ $\Delta \chi^2_{null}$}
		} 
		\startdata
	 & Q & 1.18 $\pm$ 0.35 & 3.3 & 6.09 & 7 & 0.26 & 11.17 \\
	Galaxy sample & V & 1.19 $\pm$ 0.36 & 3.3 & 6.40 & 7 & 0.32 & 11.31 \\
	 & W & 1.17 $\pm$ 0.36 & 3.3 & 6.52 & 7 & 0.21 & 10.58 \\ \\
 
	 & Q & 0.65 $\pm$ 0.74 & 0.9 & 9.12 & 7 & 0.21 & 0.77 \\
	AGN sample & V & 0.62 $\pm$ 0.74 & 0.8 & 9.86 & 7 & 0.28 & 0.70 \\
	 & W & 0.65 $\pm$ 0.74 & 0.9 & 9.32 & 7 & 0.22 & 0.79 
		\enddata
		\tablecomments{The ``d.o.f." column refers to the degrees of freedom of the $\chi^2$-distribution. $\Delta \chi^2_{null}$ shows $\Delta \chi^2$ of the best fit from the null hypothesis $\bm{t} = \bm{0}$ and $\Delta 	\chi^2_{\Lambda CDM}$ shows $\Delta \chi^2$ of the best fit from the $\Lambda$CDM prediction.}
	\end{deluxetable*}

We obtained the amplitude $A$ of the signal by minimizing $\chi^2 = (\bm{d} - A \bm{t})^T \mathrm{C}^{-1}(\bm{d} - A \bm{t})$, where $\bm{d}$ is the vector containing the measured bandpowers, $\bm{t}$ is the vector containing corresponding bandpowers of the theoretically predicted power spectra for the $\Lambda$CDM model, and $\mathrm{C}$ is the Monte Carlo covariance matrix. Then, the signal amplitude and its error are given by
\begin{equation}
	\begin{aligned}
	A &= \bm{d}^T \mathrm{C}^{-1} \bm{t} \left[ \bm{t}^T \mathrm{C}^{-1} \bm{t}\right]^{-1}, \\
	\sigma_A &= \left[ \bm{t}^T \mathrm{C}^{-1} \bm{t}\right]^{-1/2} .
	\end{aligned}
	\end{equation}
We calculated the significance of the detection from 
\begin{equation}
	\begin{aligned}
	\textrm{S/N} &= \sqrt{\chi^2_{null} - \chi^2_{fit}} \\
	&= \bm{d}^T \mathrm{C}^{-1} \bm{t} \left[ \bm{t}^T \mathrm{C}^{-1} \bm{t}\right]^{-1/2} = \frac{A}{\sigma_A} , 
	\end{aligned}
\end{equation}
where $\chi^2_{fit}$ is for the best fit and $\chi^2_{null}$ is for the null hypothesis with $\bm{t} = \bm{0}$.

We detected the ISW effect signal for AllWISE galaxies with $3.3\sigma$ significance for all three \textit{WMAP} bands. The combined ISW effect signal amplitude for the three \textit{WMAP} bands is $A = 1.18 \pm 0.36$, which agrees very well with the $\Lambda$CDM prediction of $A=1$. For AGNs, the ISW effect amplitude is $A = 0.64 \pm 0.74$ with $0.9\sigma$ significance, which is also in agreement with the $\Lambda$CDM model. The signal amplitude and some basic statistical properties for each \textit{WMAP} band are given in \autoref{table:gal-isw}.

\section{Discussion and Conclusions} \label{sect:concl}
In this study, we detected the ISW effect signal from the cross-correlation between the \textit{WMAP} CMB temperature map and the matter overdensity map using AllWISE galaxies and AGNs as tracers for matter distribution. The ISW effect detection significances for galaxies and AGNs are $3.3\sigma$ and $0.9\sigma$ respectively with a combined significance of $3.4\sigma$, with good agreement to the $\Lambda$CDM model for both tracers.

Among other ISW effect studies using \textit{WISE} data, \citet{Goto12} detected the ISW effect amplitude to be $2.2\sigma$ higher than that for the $\Lambda$CDM model, where these authors used \textit{WISE} preliminary release and \textit{WMAP} 7-year data. \citet{Ferraro15} used \textit{WISE} all-sky release and \textit{WMAP} 9-year data to detect the ISW effect signal at $3\sigma$ and in good agreement with the $\Lambda$CDM cosmology. Our result fully agrees with the finding of \citet{Ferraro15}.

The measured biases of the tracers in our study for constant and linear redshift evolution bias models are lower than those calculated by \citet{Ferraro15} by approximately 13-20\%. \citet{Ferraro15} used the lensing potential map from \textit{Planck} data release 1 (2013), whereas we used the lensing convergence map provided by \textit{Planck} data release 2 (2015). The 2013 lensing potential map was obtained by combining only the 143 and 247 GHz channels, whereas the 2015 lensing convergence map was constructed by applying a quadratic estimator to all nine frequency bands. \citet{Kuntz15} used both of the \textit{Planck} data releases to measure the cross-correlation between CMB lensing and Canada-France-Hawaii Telescope Lensing Survey (CFHTLenS) galaxy catalog and found that the cross-correlation amplitude measured using the 2015 data is roughly 19\% lower than that measured using the 2013 data. This  result is consistent with the discrepancy in the bias measurement between \citet{Ferraro15}'s and our studies.

The redshift distribution of the AllWISE galaxies might have missed a large fraction ($\sim$70\%) at the higher redshift end of the distribution due to the shallower depth of SDSS galaxies. However, this missing fraction does not significantly effect our final amplitude measurement. We checked the robustness of our measurement against errors in redshift distribution estimation by using the redshift distribution of W1 selected galaxies from \textit{WISE} all-sky release given by \citet{Yan13}  (as used by \citet{Ferraro15}) instead of our own estimation. This distribution spans a wide range of redshift up to $z\sim0.9$. We found the ISW effect amplitude for the galaxy sample to be $A=1.28 \pm 0.39$ for this galaxy redshift distribution, which is very close (within $0.3\sigma$) to the original measurement.

Contamination due to foreground emission in the CMB maps might lead to systematic error in the ISW effect detection in the form of spurious correlation with LSS tracers. However, the amount of foreground contamination would be different across the frequency bands. In our measurement, we find the cross-correlations between the LSS tracers and the CMB maps in three \textit{WMAP} bands to be consistent with each other. This consistency rules out any significant contamination by foreground emission in the CMB maps.

The significance of the ISW effect signal amplitude for our AGN sample is low ($0.9\sigma$). This low significance is partially because the AGN sample mostly spans redshift range $z \geq 1$ where the universe is not yet dominated by dark energy. As a result, the ISW effect is less sensitive to this redshift range and the expected signal becomes low. On the other hand, due to the much smaller sample size of the AllWISE AGNs, the shot noise is higher than that for the galaxy sample. This high shot noise limits the detection significance, especially in higher multipoles.

Dark energy is one of the most active fields in modern cosmology as many of its properties still remain unknown. Although the existence of dark energy is highly evidenced by various indirect measurements, the ISW effect is one of the only few direct observational probes to study dark energy. In this study, we detected the ISW effect signal by cross-correlating \textit{WMAP} CMB temperature maps with AllWISE galaxies and AGNs. These detections rule out a matter-dominated, dark-energy-free universe by a combined significance of $3.4\sigma$. Future surveys, covering a large portion of the sky with extensive redshift coverage and sufficient number of frequency bands for photometric redshift estimation, can push this detection significance to $5\sigma$ level and attain the precision necessary to pinpoint the physical properties of dark energy.

\acknowledgements We thank Alice E. Shapley, Xinnan Du, Daniel Cohen, Emily Martin, and Anson Lam for valuable comments to make this paper clearer. AJS and ELW were partially supported by the National Aeronautics and Space Administration (NASA) grant 4-443820-WR-79063-02. This publication makes use of data products from the \textit{WISE}, which is a joint project of the University of California, Los Angeles, and the Jet Propulsion Laboratory (JPL), California Institute of Technology, and NEOWISE, which is a project of the JPL, California Institute of Technology. \textit{WISE} and NEOWISE are funded by the NASA. This research made use of NASA's Astrophysics Data System; the NASA/IPAC Infrared Science Archive, which is operated by the JPL, California Institute of Technology, under contract with the NASA; matplotlib, a Python library for publication quality graphics \citep{Hunter:2007}; SciPy \citep{jones_scipy_2001}. We acknowledge the use of the Legacy Archive for Microwave Background Data Analysis (LAMBDA), part of the High Energy Astrophysics Science Archive Center (HEASARC). HEASARC/LAMBDA is a service of the Astrophysics Science Division at the NASA Goddard Space Flight Center. Funding for SDSS-III has been provided by the Alfred P. Sloan Foundation, the Participating Institutions, the National Science Foundation, and the U.S. Department of Energy Office of Science. The SDSS-III web site is \url{http://www.sdss3.org}. SDSS-III is managed by the Astrophysical Research Consortium for the Participating Institutions of the SDSS-III Collaboration.

\bibliographystyle{hapj}
\bibliography{myrefs}

\end{document}